\documentclass[a4paper]{article}
\newcommand{\be}{\begin{equation}}
\newcommand{\ee}{\end{equation}}

\newcommand{\al}{\alpha}

\newcommand{\bet}{\beta}

\newcommand{\partialp}{{\cal D}}

\newcommand{\dz}{\wedge}

\newcommand{\ba}{\begin{array}}
\newcommand{\ea}{\end{array}}
\newcommand{\beq}{\begin{eqnarray}}
\newcommand{\eeq}{\end{eqnarray}}
\newtheorem{lm}{Lemma}
\newtheorem{thee}{Theorem}
\newtheorem{proo}{Proposition}
\newtheorem{co}{Corollary}
\newtheorem{rem}{Remark}
\newtheorem{deff}{Definition}
\newcommand{\bd}{\begin{deff}}
\newcommand{\ed}{\end{deff}}

\newcommand{\bl}{\begin{lm}}
\newcommand{\el}{\end{lm}}
\newcommand{\bp}{\begin{proo}}
\newcommand{\ep}{\end{proo}}
\newcommand{\bt}{\begin{thee}}
\newcommand{\et}{\end{thee}}
\newcommand{\bc}{\begin{co}}
\newcommand{\ec}{\end{co}}
\newcommand{\brm}{\begin{rem}}
\newcommand{\erm}{\end{rem}}

\newcommand{\der}{{\rm d}}
\begin{document}

\thispagestyle{empty}

\title {Non-vacuum twisting type N metrics
\footnote{Research supported 
by Komitet Bada\'n   Naukowych (Grant nr 2 P03B 060 17)}\\
\vskip 1.truecm
{\small  Pawe\l ~Nurowski}
\footnote{{\small{\small {\tt e-mail: nurowski@fuw.edu.pl}}}}\\
\vskip -0.3truecm
{\small {\it Instytut Fizyki Teoretycznej, Uniwersytet Warszawski}}\\
\vskip -0.3truecm
{\small {\it ul. Ho\.za 69, 00-618 Warszawa, Poland}}\\
\vskip 0.3truecm
{\small  Jerzy F. Pleba\'nski}
\footnote{{\small{\small {\tt e-mail: Jerzy.Plebanski@fis.cinvestav.mx }}}} \\
\vskip -0.3truecm
{\small {\it Departamento de Fisica, CINVESTAV}}\\
\vskip -0.3truecm
{\small {\it Apdo postal 14-740, 07000 Mexico, DF}}\\
}
%\vskip -0.3truecm

\author{\mbox{}}
\maketitle
\begin{abstract}
A maximally reduced system of equations corresponding to the twisting type N 
Einstein metrics is given. When the cosmological constant $\lambda\to 0$ they 
reduce to the standard equations for the vacuum twisting type N's.  
All the metrics which are conformally equivalent to the 
twisting type N metrics and which admit 3-dimensional conformal group of
symmetries are presented. In the Feferman class of metrics an example is given of
a twisting type N metric which satisfies Bach's equations 
but is not Einstein.
\end{abstract}

\newpage
\noindent

%\vglue 1.5truecm
\rm
\section{Introduction}
A way of obtaining the maximally reduced system of equations 
corresponding to the metrics with the energy momentum tensor of the 
form $T_{\mu\nu}=k_\mu k_\nu$, where $k_\mu$ is a quadruple principal null 
direction, was presented by one of us (JFP) in an unpublished paper 
\cite{JF}. The method\footnote{The results of Ref. \cite{JF} are described in \cite{Kramer} 
pp. 240-242.} of Ref. \cite{JF} can be also applied to the vacuum 
type N equations 
\be
R_{\mu\nu}=\lambda g_{\mu\nu}\label{eni}
\ee 
with cosmological constant 
$\lambda$. It turns out that any type N metric satisfying the Einstein 
equations (\ref{eni}) is generated by a single complex function $L=L(u,z,\bar{z})$ of 
variables $u$ (real) and $z$ (complex), subjected to the following equations 
\beq
\bar{\partialp}^2\partialp
L-\partialp^2\bar{\partialp}\bar{L}=\frac{\lambda}{3}(\partialp\bar{L}-\bar{\partialp}L)^3\label{sys1}\\
\bar{\partialp}\partial_u\partialp
L=\frac{\lambda}{2}(\partialp\bar{L}-
\bar{\partialp}L)[\partialp^2\bar{L}-\bar{\partialp}\partialp L].\nonumber
\eeq  
Here, 
$\partialp=\partial_z-L\partial_u$. In terms of $L$ the metric reads
$$g=2(\theta^1\theta^2-\theta^3\theta^4)$$ 
\beq
&\theta^3=\overline{\theta^3}=\der u+L\der
z+\bar{L}\der\bar{z}\nonumber\\
&\label{met1}\\
&\theta^1=\overline{\theta^2}=[r+\frac{1}{2}(\partialp\bar{L}-\bar{\partialp}L)]\der
z+\bar{L}_u\theta^3\nonumber\\
&\nonumber\\
&\theta^4=\overline{\theta^4}=\der r+
[L_u(\partialp\bar{L}-\bar{\partialp}L)-\frac{1}{2}\partialp
(\partialp\bar{L}-\bar{\partialp}L)]\der z+
[\bar{L}_u(\bar{\partialp}L-\partialp\bar{L})-\frac{1}{2}\bar{\partialp}
(\bar{\partialp}L-\partialp\bar{L})]\der
\bar{z}+\nonumber\\
&[\frac{\lambda}{6}(-r^2+\frac{5}{4}(\partialp\bar{L}-\bar{\partialp}L)^2)-\frac{1}{2}\partial_u(\partialp\bar{L}+\bar{\partialp}L)]\theta^3,\nonumber
\eeq
and the 4-dimensional space-time is parametrized by $(r,u,z,\bar{z})$,
$r$ beeing a real coordinate. The only nonvanishing coefficient of the
Weyl tensor is 
\be
\Psi_4=\frac{\partialp^2_u\bar{\partialp}\bar{L}+\frac{\lambda}{6}(3\bar{L}_u-\bar{\partialp})(\bar{\partialp}^2L-\partialp\bar{\partialp}\bar{L})}{r-\frac{1}{2}(\partialp\bar{L}-\bar{\partialp}L)}.
\ee
The metric (\ref{met1}) has a quadruple principall null direction
$k=\partial_r$. It is twisting iff and only iff 
\be
\partialp\bar{L}-\bar{\partialp}L\neq 0.\label{twist}
\ee
If $k$ is non-twisting all the metrics corresponding to solutions of
(\ref{sys1}) are known \cite{PlebGar}. On the other hand, if condition
(\ref{twist}) is satsified no explicit solution to the equations
(\ref{sys1}) is avaliable. Even the Hauser explicit solution
\cite{Hauser} is not easily expressible in terms of function $L$
only.\\

\section{The Feferman class}
In this paper we relax the Einstein condition
and search for twisting type N metrics, which do not satisfy any
additional curvature conditions. One class of such metrics is
given by the Feferman conformal class \cite{Fef}, which in the context of GR was
first studied by Sparling \cite{Spar}. To describe the Feferman
metrics one needs the notion of a Cauchy-Riemann structure. 
\bd
A Cauchy-Riemann (CR) structure $({\cal N},[(\Omega,\Omega_1)])$ is a
3-dimensional manifold $\cal N$ equipped with a class of pairs of 
1-forms $[(\Omega,\Omega_1)]$ such that $\Omega$ is real- and $\Omega_1$ is 
complex-valued, $\Omega\dz\Omega_1\dz\bar{\Omega}_1\neq 0$ at each
point of $\cal N$, two pairs $(\Omega,\Omega_1)$ and
$(\Omega',\Omega_1')$ are equivalent iff there existsnonvanishing
functions $f$ (real) and $h$ (complex) and a complex function $p$ on
$\cal N$ such that 
\be
\Omega'=f\Omega\quad\quad\quad\quad\Omega_1'=h\Omega_1+p\Omega.\label{tr}
\ee
A Cauchy-Riemann structure is nondegenerate iff
$\der\Omega\dz\Omega\neq 0$.
\ed 
Given a nondegenerate CR structure $({\cal N},[(\Omega,\Omega_1)])$ one can always choose a
representative $(\Omega,\Omega_1)$  from the class
$[(\Omega,\Omega_1)]$ such that \cite{Cart,phd}
\be
\der\Omega=i\Omega_1\dz\bar{\Omega}_1.\label{car}
\ee
Since $(\Omega,\Omega_1,\bar{\Omega}_1)$ constitutes a basis of
1-forms on $\cal N$ then the differential of $\Omega_1$ uniquely
defines functions $\al,\theta$ (complex) and $\beta$ (real)
such that 
\be
\der\Omega_1=\bar{\al}\Omega_1\dz\bar{\Omega}_1+i\bet\Omega\dz\Omega_1-
\theta\Omega\dz\bar{\Omega}_1.\label{ala}
\ee
Let $(\partial_0,\partial,\bar{\partial})$ be a basis of vector fields on
$\cal N$ dual to $(\Omega,\Omega_1,\bar{\Omega}_1)$,
respectively. Then the equation $\der^2\Omega_1\equiv 0$ implies the
following identity
\be
\partial_0\al-i\partial\bet+\bar{\partial}\bar{\theta}+i\bet\al-\bar{\al}\bar{\theta}=0.\label{id}
\ee
It is convenient to introduce the following operators
\be
\triangle=\partial-\al\quad\quad\quad\quad\delta=\partial_0+i\bet.
\ee
Condition (\ref{tr}) does not fix the forms $(\Omega,\Omega_1)$
totally. They still can be transformed by means of the following
transformations 
\be
\Omega\to t\bar{t}\Omega\quad\quad\quad\quad
\Omega_1\to t[\Omega_1+i\bar{\partial}\log(t\bar{t})\Omega],\label{tr1}
\ee 
where $t$ is a nonvanishing complex function on $\cal N$.
The corresponding transformations of functions $\al,\bet,\theta$ are
\beq
&\al\to\frac{1}{t}(\al-\partial\log(t\bar{t}^2))\nonumber\\
&\theta\to\frac{1}{\bar{t}^2}(\theta+\bar{\al}z+\bar{\partial}z+iz^2)\\
&\bet\to\frac{1}{|t|^2}[\bet-\bar{\partial}\partial\log t-\partial\bar{\partial}\log\bar{t}-\al\bar{\partial}\log
t-\bar{\al}\partial\log\bar{t}2(\bar{\partial}\log
t\partial+\partial\log\bar{t}\bar{\partial})\log |t|],\nonumber
\eeq
where $z=i\bar{\partial}\log(t\bar{t})$.\\

\noindent
Let $({\cal N},[(\Omega,\Omega_1)])$ be a CR-structure with
$(\Omega,\Omega_1)$ satisfying (\ref{car}). One defines a manifold
\be
{\cal M}={\bf R}\times\cal N\label{st}
\ee 
with a canonical projection 
$\pi:{\cal M}\to\cal N$ and pull-backs the forms $(\Omega,\Omega_1)$
to $\cal M$ by means of $\pi$. Then, using the same letters to denote
the pull-backs, one equipps $\cal M$ with a class of Lorentzian
metrics of the form
\be
g={\rm e}^{2\phi}[\Omega_1\bar{\Omega}_1-\Omega (\der
r+W\Omega_1+\bar{W}\bar{\Omega}_1+H\Omega)].
\label{met}
\ee
Here $\phi, H$ (real) and $W$ (complex) are arbitrary functions on
$\cal N$ and $r$ is a real coordinate along the factor ${\bf R}$ in
${\cal M}={\bf R}\times\cal N$.\\

\noindent
It follows that the vector field $k=\partial_r$ is null geodesic and
shear-free in any metric (\ref{met}). It generates a congruence of
shear-free and null geodesics in $\cal M$ which is always twisting due
to condition (\ref{car}). The converse is also true. Any space-time
admitting a twisting congruence of shear-free and null geodesics can
be obtained in this way \cite{trautopt}. This, in particular, means
that any such sace-time defines its corresponding CR-structure - the
3-dimensional manifold of the lines of the congruence.\\

\noindent
In the context of the present paper it is 
interesting to ask when the metrics (\ref{met}) are of type N with $k$
beeing a quadruple principal null direction. The answer is given by
the following theorem \cite{msc,lewnur}.
\bt
The metric (\ref{met}) has $k=\partial_r$ as a quadruple principal
null direction if and only if
\beq
&W=2ai{\rm e}^{ir}+b\nonumber\\
&H={\rm e}^{ir}(\bar{\triangle}-i\bar{b})a+{\rm
e}^{-ir}(\triangle+ib)\bar{a}+h\label{pic}\\
&h=i(\triangle\bar{b}-\bar{\triangle}b)-6a\bar{a}-\frac{1}{2}(\triangle\bar{\al}+\bar{\triangle}\al+\bet)\nonumber
\eeq
where the complex functions $a:{\cal N}\to{\bf C}$ and $b:{\cal N}\to
{\bf C}$ satisfy 
\beq
&2iha-2\delta
a-i\partial\bar{\triangle}a-\partial(\bar{b}a)-b\bar{\triangle}a+ib\bar{b}a=0\label{n1}\\
&4\bar{\al}\bar{\theta}-2\bar{\partial}\bar{\theta}+3i(\delta b-\bar{\theta}\bar{b})+\partial(\bar{\triangle}b-\triangle\bar{b}-4ih)+8i(a\triangle\bar{a}-\partial(a\bar{a})+iba\bar{a})=0.\label{n2}
\eeq 
\et 

\noindent
Take 
\be
a=0,\quad\quad\quad\quad b=\frac{2}{3}i\al.
\ee
Then equation (\ref{n1}) is automatically satisfied and equation
(\ref{n2}) becomes the identity (\ref{id}). Then, applying Theorem 1
we have the following Corollary.
\bc
Let $({\cal N},[(\Omega,\Omega_1)])$ be a CR-structure generated by
forms  $(\Omega,\Omega_1)$ satisfying condition (\ref{car}). Then the
metric 
\be
g={\rm e}^{2\phi}\{\Omega_1\bar{\Omega}_1-\Omega [\der
r+\frac{2}{3}i\al\Omega_1-\frac{2}{3}i\bar{\al}\bar{\Omega}_1+\frac{1}{6}(\triangle\bar{\al}+\bar{\triangle}\al-3\bet)\Omega]\}\label{fef}
\ee
on manifold ${\cal M}={\bf R}\times\cal N$ is of type N with twisting
shear-free null geodesics generated by the quadruple principal null dierction $k=\partial_r$.
\ec
The metrics (\ref{fef}) are called the Feferman metrics
\cite{Fef}. Their main property is presented in the following
theorem.
\bt~\\
Let a pair $(\Omega,\Omega_1)$ satisfying (\ref{car}) undergoes 
transformation (\ref{tr1}). Then the metric $g$ of (\ref{fef})
transforms according to\footnote{To get this result one has to
redefine the $r$ coordinate according to $r\to r+\frac{i}{3}\log\frac{t}{\bar{t}}$}
$g\to t\bar{t}g$. 
\et
Thus, any nondegenerate CR-structure defines a conformal class of
Feferman metrics (\ref{fef}). Each of the metrics in the class is of
type N and its quadruple principal null direction defines a congruence
of shear-free and null geodesics with twisting rays. We stress here
that Corollary 1 provides an effective method of evaluating Feferman twisting
type N metrics for each nondegenrate CR-structure. For example, if the
CR-structure $\cal N$ is embeddable in ${\bf C}^2$ (cf. \cite{traut},
p. 499 for
definition of embeddability) it may be parametrized by coordinates
$(u,z,\bar{z})$, $u$ - real, $z$ - complex, and generated by a free complex
function $L=L(u,z,\bar{z})$ such that
$\partialp\bar{L}-\bar{\partialp}L\neq 0$. 
The 1-forms $(\Omega,\Omega_1)$ satisfying (\ref{car})
may be choosen to be 
\be
\Omega=i\frac{\der u+L\der z+\bar{L}\der\bar{z}}{\partialp\bar{L}-\bar{\partialp}L}
\ee
$$
\Omega_1=\der z-i\bar{w}\Omega,\label{eff}
%[\bar{L}_u+\bar{\partialp}\log(\partialp\bar{L}-\bar{\partialp}L)]\Omega.
$$
where 
\be
w=L_u+\partialp\log(\partialp\bar{L}-\bar{\partialp}L).\label{ew}
\ee
Then, the Feferman metric (\ref{fef}) is 
\be
g={\rm e}^{2\phi}\{\der z\der\bar{z}-\Omega [\der
r-\frac{i}{3}w\der z+\frac{i}{3}\bar{w}\der\bar{z}-\frac{1}{6}
\partialp\bar{w}\Omega]\}\label{fefemb}
\ee
%\beq
%&\al=w\nonumber\\
%L_u+\partialp\log(\partialp\bar{L}-\bar{\partialp}L)\nonumber\\
%&\bet=\partialp
%[\bar{L}_u+\bar{\partialp}\log(\partialp\bar{L}-\bar{\partialp}L)]\nonumber\\
%&\theta=-i\bar{\partialp}[\bar{L}_u+\bar{\partialp}\log(\partialp\bar{L}-\bar{\partialp}L)]\label{ef}\\
%&\partial=\partialp\nonumber\\
%&\partial_0=i(\bar{\partialp}L-\partialp\bar{L})\partial_u+i[\bar{L}_u+\bar{\partialp}\log(\partialp\bar{L}-\bar{\partialp}L)]\partialp-i[L_u+\partialp\log(\partialp\bar{L}-\bar{\partialp}L)]\bar{\partialp}\nonumber.
%\eeq
Inserting (\ref{eff})-(\ref{ew}) into (\ref{fefemb}) gives an explicit form of the
Feferman metric for each embeddable CR-structure.\\

\noindent
Another characterization of the Feferman class of metrics is given by
the following theorem \cite{Spar,lewphd}.
%\newpage
\bt
Feferman metrics $g$ are the only metrics satisfying the following three conditions:
\begin{itemize}
\item[(a)] $g$ are of type N with a quadruple principal null direction
generated by a vector filed $k$,
\item[(b)] k is geodesic, shear-free and twisting,
\item[(c)] k is a conformal Killing vector field.
\end{itemize}
\et

\noindent
A bit disapointing property of the Feferman class is given below
\cite{lew}.

\bt
None of the Feferman metrics satisfies Einstein equations
$R_{\mu\nu}=\lambda g_{\mu\nu}$.
\et

\section{Twisting type N metrics with 3-dimensional group of conformal 
symmetries}
In this section we find a local form of all Lorentzian metrics $g$ 
which satisfy the following assumptions
\begin{itemize}
\item[(i)] $g$ are of type N with $k$ beeing a quadruple principal null 
direction,
\item[(ii)] $k$ is geodesic, shear-free and twisting,
\item[(iii)] $g$ is not conformally equivalent to any of the Feferman metrics 
and not conformally flat,
\item[(iv)] $g$ admit at least 3 conformal Killing vector fields.
\end{itemize}
The following theorem is implicit in Sections 4-6 of
Ref. \cite{lewnur}.
\bt
All the metrics satisfying assumptions (i)-(iv) can locally be
represnted by 
\be
g={\rm e}^{2\phi}[\Omega_1\bar{\Omega}_1-\Omega (\der
r+W\Omega_1+\bar{W}\bar{\Omega}_1+H\Omega)],
\ee
where 
\be
W=2ai{\rm e}^{ir}+b,\quad\quad\quad H=-a[{\rm
e}^{ir}(\bar{\al}+i\bar{b})+{\rm e}^{-ir}(\al-ib)]+
i(\bar{\al}b-\al\bar{b})-6a^2-\frac{1}{2}\bet+\al\bar{\al}
\ee
and the functions $a>0,b,\al,\bet,\theta$ are
constants \footnote{If $a\equiv 0$ then the corresponding metric 
is in the Feferman class.}, satisfying
\be
-12i a^2 - \bar{\al} b + i b \bar{b} + 
  2 \al (i\bar{\al} + \bar{b}) - 3 i\bet=0\label{n1c}
\ee
\be
-8 i a^2 (\al - i b) - 3 b \bet + 4 \bar{\al}\bar{\theta} - 
  3 i\bar{b} \bar{\theta}=0.\label{n2c}
\ee
$\Omega$ and $\Omega_1$ are related to $\al,\bet,\theta$ by
(\ref{ala}) 
and satisfy (\ref{car}). The space-time is locally a cartesian product
${\cal M}={\bf R}\times\cal N$, with $({\cal N},(\Omega,\Omega_1))$ 
beeing a three dimensional nondegenerate CR-structure. 
The coordinate $r$ is choosen so that the orbits of the three 
conformal symmetries $X_i$,
$i=1,2,3$ are given by $r$=const and $X_i(r)=0$. The three symmetries $X_i$ 
are such that 
\be
{\cal L}_{X_i}\Omega={\cal L}_{X_i}\Omega_1=0\quad\quad\quad\forall i=1,2,3,
\ee 
so that they constitute also three symmetries of the CR-structure 
$({\cal N},(\Omega,\Omega_1))$.
\et
It follows from this theorem that all the metrics satisfying (i)-(iv)
can be obtained by inspecting the list \cite{Cart,nurtaf,lewnur} 
of all nondegenerate CR-structures admitting three symmetries. Such
structures are classified according to the Bianchi type of the
corresponding symmetries. For each Bianchi type the forms $\Omega,\Omega_1$
and the constants $\al,\bet,\theta$ are presented in
Ref. \cite{lewnur}. Using this list, one has to check whether a given
Bianchi type represented by constants $\al,\bet,\theta$ is admitted by
the type N equations (\ref{n1c})-(\ref{n2c}). If it is, one finds the
corresponding $a$ and $b$.\\

\noindent
It turns out that only CR-structures with symmetry groups of Bianchi
types $VI_h$ and $VIII$ are admitted by equations
(\ref{n1c})-(\ref{n2c}). Below we describe the corresponding
solutions.
\subsection{Solutions for Bianchi type $VIII$}
In this case one has a 1-parameter family of nonequivalent CR-structures, 
parametrized by $k\geq 0$, $k\neq 1$. The manifold $\cal N$ of such 
CR-structures can be coordinatized by $(u,z,\bar{z})$, 
($u$-real, $z$-complex) and the forms 
$\Omega,\Omega_1$ and the constants $\al,\bet,\theta$ of Theorem 5 can be 
choosen so that
\beq
&\Omega=\frac{2}{k^2-1}\lambda_0\quad\quad\quad\Omega_1=\mu_0-\frac{k}{k^2-1}\lambda_0\nonumber\\
&\nonumber\\
&\mu_0=\frac{2{\rm e}^{iu}}{z\bar{z}-1}\der z
\quad\quad\quad\lambda_0=\der u+\frac{k{\rm e}^{iu}-i\bar{z}}
{z\bar{z}-1}\der z+\frac{k{\rm e}^{-iu}+iz}{z\bar{z}-1}\der \bar{z}\label{8}\\
&\nonumber\\
&\al=0\quad\quad\quad\bet=\frac{1}{4}(k^2-2)\quad\quad\quad\theta=-\frac{i}{4}k^2.\nonumber
\eeq
Inserting the above $\al,\bet,\theta$ to the type N equations (\ref{n1c})-(\ref{n2c}) one finds two branches of solutions for $a$ and $b$. The corresponding metrics are 
\be
g={\rm e}^{2\phi}[\Omega_1\bar{\Omega}_1-\Omega \nu]
\ee 
where the forms $\Omega$ and $\Omega_1$ are given by (\ref{8}) and the real 1-form $\nu$ is given below for each branch (a) and (b) separately.
\begin{itemize}
\item[(a)] If $0\leq k\leq\frac{\sqrt{2}}{2}$ then 
\beq
&\nu=\der r+\frac{i}{2}\sqrt{3(1-k^2)}
({\rm e}^{ir}\pm\sqrt{\frac{1-2k^2}{1-k^2}})\Omega_1-\frac{i}{2}\sqrt{3(1-k^2)}({\rm e}^{-ir}\pm\sqrt{\frac{1-2k^2}{1-k^2}})\bar{\Omega}_1+\nonumber\\
&(\mp\frac{3}{8}\sqrt{(1-2k^2)(1-k^2)}({\rm e}^{ir}+{\rm e}^{-ir})+k^2-\frac{7}{8})\Omega.\nonumber
\eeq
Solutions are not conformally flat iff $k\neq\frac{\sqrt{2}}{2}$. 
\item[(b)] If $k\geq 0$, $k\neq 1$, $k\neq\sqrt{3}$ then
\beq
&\nu=\der r+\frac{\sqrt{3}}{2}(i{\rm e}^{ir}\pm\sqrt{1+k^2})\Omega_1+
\frac{\sqrt{3}}{2}(-i{\rm e}^{-ir}\pm\sqrt{1+k^2})\bar{\Omega}_1+\nonumber\\
&\frac{1}{8}(\mp 3i\sqrt{1+k^2}({\rm e}^{ir}-{\rm e}^{-ir})-k^2-7)\Omega.\nonumber
\eeq
Solutions are not conformally flat iff $k\neq 1$, $k\neq\sqrt{3}$.
\end{itemize} 
\subsection{Solutions for Bianchi type $VI_h$} 
In this case one has only one CR-structures for each value of the real 
parameter $h=-(\frac{1-d}{1+d})^2$, $-1<d\leq1$.  For each value of $d$ 
the CR-manifold $\cal N$ can be coordinatized by real $(u,x,y)$, and the forms 
$\Omega,\Omega_1$ and the constants $\al,\bet,\theta$ of Theorem 5 can be 
choosen so that
\beq
&\Omega=-\frac{2}{d+1}\lambda_0\quad\quad\quad\Omega_1=\mu_0+
\frac{d}{d+1}\lambda_0\nonumber\\
&\nonumber\\
&\mu_0=y^{-1}\der (x+iy)
\quad\quad\quad\lambda_0=y^d\der u-y^{-1}\der x\label{6h}
&\nonumber\\
&\al=\frac{i}{2}(d-1)\quad\quad\quad\bet=-\frac{1}{4}d\quad\quad\quad\theta=
-\frac{i}{4}d.\nonumber
\eeq
For the above $\al,\bet,\theta$ the type N equations (\ref{n1c})-(\ref{n2c}) 
imply that the constant $b$ is real and satsifies 
\be
8b^3+8(d-1)b^2-2(1+4d+d^2)b-2d^3-3d^2+3d+2=0.\label{bd}
\ee 
Note that equation (\ref{bd}) always admits at least one real solution. 
Once the real solution $b=b(d)$ of this equation is known one has to check, 
whether the quantity 
$$
A=\frac{1}{48}[4b^2+6(d-1)b+2d^2-d+2]
$$ 
is positive. If it is positive, the constant 
\be
a=\sqrt{A}.\label{bd1}
\ee 
Otherwise there is no 
solution to (\ref{n1c})-(\ref{n2c}) corresponding to $b=b(d)$. \\
To describe an example of explicit solutions to equations 
(\ref{bd})-(\ref{bd1}) we choose 
\be
d=\frac{1}{2}.
\ee 
Then, there are three 
different solutions for $a$ and $b$:
\be
b_1=-\frac{1}{4}(\sqrt{6}+1)\quad\quad\quad a_1=\frac{1}{8}\sqrt{6+\frac{5}{3}\sqrt{6}},
\ee
\be
b_2=1\quad\quad\quad a_2=\frac{1}{4}, 
\ee
\be
b_3=\frac{1}{4}(\sqrt{6}-1)\quad\quad\quad a_3=\frac{1}{8}\sqrt{6-\frac{5}{3}\sqrt{6}}.
\ee
The corresponding metrics are given by 
\be
g={\rm e}^{2\phi}[\Omega_1\bar{\Omega}_1-\Omega \nu]
\ee 
where the forms $\Omega$ and $\Omega_1$ are given by (\ref{6h}) with 
$d=\frac{1}{2}$, and the real 1-form $\nu$ is given below for each of the 
three above solutions by
\beq
&\nu_1=\der r+\frac{1}{4}[i\sqrt{6+\frac{5}{6}\sqrt{6}}{\rm e}^{ir}-\sqrt{6}-1]\Omega_1+\frac{1}{4}[-i\sqrt{6+\frac{5}{6}\sqrt{6}}{\rm e}^{-ir}-\sqrt{6}-1]\bar{\Omega}_1+\nonumber\\
&\frac{1}{32}[i\sqrt{36+10\sqrt{6}}({\rm e}^{ir}-{\rm e}^{-ir})-\sqrt{6}-10]\Omega\nonumber
\eeq 
\beq
&\nu_2=\der r+[\frac{i}{2}{\rm e}^{ir}+1]\Omega_1+[-\frac{i}{2}{\rm e}^{-ir}+1]\bar{\Omega}_1+\nonumber\\
&[\frac{5}{16}i({\rm e}^{-ir}-{\rm e}^{ir})-\frac{3}{4}]\Omega\nonumber
\eeq 
\beq
&\nu_3=\der r+\frac{1}{4}[i\sqrt{6-\frac{5}{6}\sqrt{6}}{\rm e}^{ir}+\sqrt{6}-1]\Omega_1+\frac{1}{4}[-i\sqrt{6-\frac{5}{6}\sqrt{6}}{\rm e}^{-ir}+\sqrt{6}-1]\bar{\Omega}_1+\nonumber\\
&\frac{1}{32}[i\sqrt{36-10\sqrt{6}}({\rm e}^{-ir}-{\rm e}^{ir})+\sqrt{6}-10]\Omega\nonumber
\eeq 
Each of these metrics is conformally non-flat.\\

\noindent
Although equation (\ref{bd}) can be solved explicitely, the formula for $b$ 
is not very useful in obtaining the explicit forms of the metrics. It is more 
convenient to solve equation (\ref{bd}) for particular values of $d$ as we 
did above for $d=\frac{1}{2}$. Instead of giving further examples we present 
the qualitative description of the solutions, which was obtained by 
numerical analysis of (\ref{bd})- (\ref{bd1}). We have three possible 
branches of solutions, corresponding to three different roots 
$b_1(d),b_2(d),b_3(d)$ of (\ref{bd}). These branches are as follows.
\begin{itemize}
\item[$b_1(d)$] The solutions corresponding to the first root
$b_1(d)$ are only possible if $d\geq -0.511878$. It turns out then,
that for each value of $d\geq -0.511878$ there exists a type N metric,
which is not conformally flat iff $d\neq -\frac{1}{2}$ and $d\neq 0$. 
\item[$b_2(d)$] The second root $b_2(d)$ admits solutions for each
value of the parameter $d\geq0$. For each such $d$ there exists
precisely one type N metric, which is non conformally flat iff $d\neq 0$.
\item[$b_3(d)$] The third root $b_3 (d)$ admits solutions only for 
$-0.511878\leq d\leq-0.220789$ or $d\geq0$. For each such $d$
there exists precisely one type N metric which is conformally
non-flat iff $d\neq -0.333347$, $d\neq -0.220789$, $d\neq 0$, $d\neq 1$.
\end{itemize}

\noindent
We close this section with a remark that the metrics
satisfying assumptions (i)-(iv) are not conformally equivalent to the Ricci
flat metrics. This result follows directly from the analysis performed
in Ref. \cite{lewnur}.

\section{Examples of twisting type N metrics admitting two conformal
symmetries.}
In this section we present examples of type N metrics admitting two conformal
symmetries. We additionally assume that the metrics are not
conformally flat and that they do not belong to the Feferman
class. The general solution for such a problem is rather hopeless to
obtain but the following two examples can be given.\\
{\bf Case A.}\\
Consider a 3-dimensional manifold parametrized by the real coordinates
$(u,x,y)$. The CR-structure $({\cal N},[(\Omega,\Omega_1)])$ is
generated on $\cal N$ by the forms 
\be
\Omega=\der u-y\der x\quad\quad\quad\Omega_1=\frac{\sqrt{2}}{2}(\der
x+i\der y).\label{oo1}
\ee 
The forms (\ref{oo1}) satisfy (\ref{car}). On ${\cal M}={\bf
R}\times\cal N$ introduce a coordinate $r$ along the ${\bf R}$ factor
and consider the metric 
\be
g={\rm e}^{2\phi}[\Omega_1\bar{\Omega}_1-\Omega \nu]\label{ca1}
\ee
with the 1-form $\nu$ defined by
\beq
&\nu=\der
r+y^{-1}[(\frac{\sqrt{2}}{2}-\sqrt{3}+(\frac{3}{2})^{\frac{1}{4}}i{\rm
e}^{ir})\Omega_1+(\frac{\sqrt{2}}{2}-\sqrt{3}-(\frac{3}{2})^{\frac{1}{4}}i{\rm
e}^{-ir})\bar{\Omega}_1]+\nonumber\\
&y^{-2}[\frac{i6^{\frac{1}{4}}}{4}(\sqrt{6}-2)({\rm e}^{ir}-{\rm e}^{-ir})+\frac{\sqrt{6}}{4}-1]\Omega\label{ca1'}
\eeq
It is a matter of straigthforward calculation to see that the so
defined metric is of type N, admits a congruence of twisting shear-free and
null geodesics aligned with the principal null direction 
and is never conformally flat. Moreover it has only two conformal
symmetries $X_1=\partial_u$, $X_2=\partial_x$.\\
{\bf Case B}\\
Now the CR-structure $({\cal N},[(\Omega,\Omega_1)])$ is
generated by the forms 
\be
\Omega=\der u-y^2\der x\quad\quad\quad\Omega_1=\sqrt{y}(\der
x+i\der y).\label{oo2}
\ee 
The forms (\ref{oo2}) satisfy (\ref{car}). The type N metric is defined on 
${\cal M}={\bf R}\times\cal N$ by 
\be
g={\rm e}^{2\phi}[\Omega_1\bar{\Omega}_1-\Omega \nu]\label{ca2}
\ee
with the 1-form $\nu$ defined by
\beq
&\nu=\der
r+\frac{1}{2y^{3/2}}[(i\sqrt{5+2\sqrt{19}}{\rm
e}^{ir}-\sqrt{19})\Omega_1+(-i\sqrt{5+2\sqrt{19}}{\rm
e}^{-ir}-\sqrt{19})\bar{\Omega}_1]+\nonumber\\
&\frac{i}{8y^3}[\sqrt{5+2\sqrt{19}}(\sqrt{19}-1)({\rm e}^{ir}-{\rm e}^{-ir})-16-2\sqrt{19}]\Omega.\label{ca2'}
\eeq
Here $r$ is a coordinate $r$ along the factor ${\bf R}$ in $\cal M$. \\
The above metric is of type N, admits a congruence of twisting shear-free and
null geodesics aligned with the principal null direction 
and is never conformally flat. It has only two conformal
symmetries $X_1=\partial_u$, $X_2=\partial_x$.
\section{Example of a type N metric with vanishing Bach tensor and not conformal to an Einstein metric}
It is interesting to ask whether metrics (\ref{ca1})-(\ref{ca1'}), 
(\ref{ca2})-(\ref{ca2'}) are conformally equivalent to Einstein metrics. It 
is known that a neccessary condition for a metric to be conformal to
an Einstein metric is that its Bach tensor 
\be
B_{\mu\nu}=C_{\mu\rho\nu\sigma}^{~~~~~;\sigma\rho}+\frac{1}{2}
C_{\mu\rho\nu\sigma}R^{\rho\sigma}
\ee
identically vanishes \cite{masbas,newm}. (We 
denoted the Weyl conformal curvature by $C_{\mu\rho\nu\sigma}$.)\\

\noindent
With a help of an extremely powerful symbolic
algebra package GRTensor \cite{grtensor} we calculated 
the Bach tensor for metrics (\ref{ca1})-(\ref{ca1'}) and  
(\ref{ca2})-(\ref{ca2'}). In both cases it is never vanishing, so we
conclude that the metrics (\ref{ca1})-(\ref{ca1'}) and  
(\ref{ca2})-(\ref{ca2'}) are not conformally equivalent to any
Einstein metric.\\

\noindent
Surprisingly within the Feferman class we found a metric which has
vanishing Bach tensor. This metric reads as follows:
\be
g={\rm e}^{2\phi}[\der x^2+\der y^2-\frac{2}{3}(\der x+y^3\der
u)(y\der r+\frac{11}{9}\der x -\frac{1}{9}y^3\der u)].
\ee
This metric, like any other from the Feferman class, is of type N,
admits a 
congruence of twisting
shear-free and null geodesics aligned with the principal null direction 
and is not conformally flat. It is interesting that it satisfies the
Bach equations and, beeing in the Feferman class, is not conformal to
any 
Einstein metrics. To the best of our knowledge this is the only known 
example of a Lorentzain metric having this last property \cite{mason}.

\end{document}